\documentstyle[preprint,pre,aps,oldlfont]{revtex}
\tightenlines
\begin{document}
\draft

\title{Bifurcations and Transitions to Chaos in An Inverted Pendulum}

\author{Sang-Yoon Kim${}^{a,b}$ 
\footnote{Electronic address: sykim@cc.kangwon.ac.kr}
 and Bambi Hu${}^{b,c}$
       }
 \address{
 ${}^a$Department of Physics, Kangwon National University,
 Chunchon, Kangwon-Do 200-701, Korea \\
 ${}^b$ Centre for Nonlinear Studies and Department of Physics, 
 Hong Kong Baptist University, Hong Kong, China \\
 ${}^c$ Department of Physics, University of Houston, Houston, 
 TX 77204
 }

\maketitle

\begin{abstract}
We consider a parametrically forced pendulum with a vertically 
oscillating suspension point. It is well known that, as the amplitude 
of the vertical oscillation is increased, its inverted state 
(corresponding to the vertically-up configuration) undergoes a cascade 
of ``resurrections,'' i.e., it becomes stabilized after its 
instability, destabilize again, and so forth {\it ad infinitum}. We 
make a detailed numerical investigation of the bifurcations associated 
with such resurrections of the inverted pendulum by varying the 
amplitude and frequency of the vertical oscillation. It is found that 
the inverted state stabilizes via alternating ``reverse'' subcritical 
pitchfork and period-doubling bifurcations, while it destabilizes via 
alternating ``normal'' supercritical period-doubling and pitchfork
bifrucations. An infinite sequence of period-doubling bifurcations, 
leading to chaos, follows each destabilization of the inverted 
state. The critical behaviors in the period-doubling cascades are 
also discussed.
\end{abstract}


%
%

\narrowtext

\section{Introduction}
\label{sec:Int}

We consider a parametrically forced pendulum whose suspension point
undergoes a vertical periodic oscillation. The system is 
described by a second-order non-autonomous ordinary differential 
equation \cite{PFP},
\begin{equation}
I {\ddot \theta} + b {\dot  \theta} + m l (g - \epsilon \omega^2
 \cos \omega t) \sin \theta = 0,
\label{eq:PFP1}
\end{equation}
where the overdot denotes the differentiation with respect to time, 
$I$ is the total moment of inertia, $b$ is a damping 
coefficient, $m$ is a mass attached to one end of a light rigid rod 
(its mass can be negligible) of length $l$, $\theta$ is the angular 
displacement measured counterclockwise from the downward
vertical, and $\epsilon$ and $\omega$ are the driving amplitude and
frequency of the vertical oscillation of the suspension point, 
respectively. Making the normalization 
$ \omega t \rightarrow 2 \pi t$ and $\theta \rightarrow 2 \pi x$, 
we obtain a dimensionless normalized form of Eq.~(\ref{eq:PFP1}),
\begin{equation}
{\ddot x} + 2 \pi \beta \Omega {\dot x} + 2 \pi
(\Omega^2 - A \cos 2 \pi t) \sin 2 \pi x = 0,
\label{eq:PFP2}
\end{equation}
where $\omega_0 = \sqrt{ {mgl} / I} $, 
$\beta = {b /{I \omega_0}}$, $\Omega = { \omega_0 / \omega}$, 
and $A = { {ml \epsilon} / I}$.

The parametrically forced pendulum has two stationary states.
One is the ``normal'' state corresponding to the vertically-down
configuration with $x=0$, and the other one is the ``inverted'' state 
corresponding to the vertically-up configuration with $x={1 \over 2}$. 
For the case of the ``unforced'' simple pendulum (with $A=0$), the 
normal state is obviously stable, while the inverted state is clearly 
unstable. However, as the normalized driving amplitude $A$ is 
increased above a critical value, the inverted state becomes stable. 
This stabilization of the inverted pendulum has been discussed 
theoretically 
\cite{Corben,Kapitza,Stoker,Landau,Arnold,Levi,Blackburn1} 
and demonstrated experimentally \cite{Friedman,Machaelis,Blackburn2}.

Here we are interested in the bifurcations associated with stability
of the inverted state. As in the case of the normal state \cite{Kim}, 
the linear stability of the inverted state is determined by a damped 
Mathieu equation \cite{Rem},
\begin{equation}
{\ddot u} + 2 \pi \beta \Omega {\dot u} + 4 \pi^2
(-\Omega^2 + A \cos 2 \pi t) u = 0.
\label{eq:DME}
\end{equation}
It is well known that the Mathieu equation has an infinity of 
alternating stable and unstable $A$ ranges for a given $\Omega$
\cite{ME}. Consequently, as the parameter $A$ is increased, 
the inverted pendulum exhibits a cascade of 
``resurrections'' (i.e., it becomes stabilized after its instability,
destabilizes again and so forth {\it ad infinitum}) for any given 
$\Omega$. By varying the normalized driving amplitude $A$ and the
normalized natural frequency $\Omega$, we make a detailed numerical 
investigation of bifurcation behaviors associated with such 
resurrections of the inverted pendulum for a fixed value of the
normalized damping coefficient $\beta$.

This paper is organized as follows. In Sec.~\ref{sec:SBLW}, we 
discuss bifurcations associated with stability of 
periodic orbits, using the Floquet theory \cite{Lefschetz1}.
The bifurcation behaviors associated with the resurrections of the 
inverted state are then investigated through numerical
calculationa of its Floquet (stability) multipliers in 
Sec.~\ref{sec:RES} for the case $\beta=0.2$. It is found that 
the stabilizations of the inverted state occur via alternating 
``reverse'' subcritical pitchfork and period-doubling bifurcations, 
while its destabilizations take place through alternating ``normal'' 
supercritical period-doubling and pitchfork bifrucations.
After each destabilization of the inverted state, an infinite 
sequence of period-doubling bifurcations follows and leads to chaos.
In Sec.~\ref{sec:CB}, we also study the critical behaviors in the 
period-doubling cascades. Finally, a summary is given in 
Sec.~\ref{sec:Sum}.

\section{Stability, Bifurcations, Lyapunov Exponents, and Winding 
         Numbers}
\label{sec:SBLW}

In this section, we first discuss stability of periodic orbits in
the Poincar\'{e} map of the parametrically forced pendulum, using 
the Floquet theory \cite{Lefschetz1}. Bifurcations associated with
the stability, Lyapunov exponents and winding numbers are 
then discussed.

The second-order ordinary differential equation (\ref{eq:PFP2}) is
reduced to two first-order ordinary differential equations:
\begin{mathletters}
\begin{eqnarray}
{\dot x} &=& y,  \\
{\dot y} &=& f(x,y,t) = -2 \pi \beta \Omega y - 2 \pi (\Omega^2 - A 
            \cos 2 \pi t)  \sin 2 \pi x.
\end{eqnarray}
\label{eq:PFP3}
\end{mathletters}
These equations have the (space) inversion symmetry $S$, because
the transformation
\begin{equation}
S: x \rightarrow -x,~y \rightarrow -y,~ t  \rightarrow t,
\label{eq:S}
\end{equation}
leaves Eq.~(\ref{eq:PFP3}) invariant.
If an orbit $z(t) [\equiv (x(y),y(t))]$ is invariant under $S$, 
it is called a symmetric orbit. Otherwise, it is called an 
asymmetric orbit and has its ``conjugate'' orbit $S z(t)$.

The surface of section for the parametrically forced pendulum
is the Poincar\'{e} (time-$1$) map. Hence the Poincar\'{e} maps of an 
initial point $z_0$ $[=(x_0,y_0)]$ can be computed by sampling the 
orbit points $z_m$ at the discrete time $t=m$ $(m=1,2,3, \dots)$. We 
call the transformation $z_m \rightarrow z_{m+1}$ the Poincar\'{e} 
map and write $z_{m+1}= P (z_m)$.

The linear stability of a $q$-periodic orbit of $P$ such that
$P^q(z_0) = z_0$ is determined from the linearized-map matrix
$DP^q$ of $P^q$ at an orbit point $z_0$. Here $P^q$ means the 
$q$-times iterated map. Using the Floquet theory \cite{Lefschetz1}, 
the matrix $DP^q$ can be obtained by integrating the linearized 
differential equations for small perturbations as follows.

Let $z^*(t)=z^*(t+q)$ be a solution lying on the closed orbit
corresponding to the $q$-periodic orbit. In order to determine the
stability of the closed orbit, we consider an infinitesimal
perturbation $\delta z [=(u,v)]$ to the closed orbit.
Linearizing Eq.~(\ref{eq:PFP3}) about the closed orbit, we obtain
\begin{equation}
 \left( \begin{array}{c}
         {\dot u}  \\
         {\dot v}
      \end{array}
      \right)
      = J(t)
      \left( \begin{array}{c}
         u  \\
         v
      \end{array}
      \right),
 \,\,\,\,\,J(t) =
 \left( \begin{array} {cc}
        0 & 1 \\
        f_x (x^*,t) & f_y
        \end{array}
 \right).
\label{eq:LE1}
\end{equation}
Here $f_x$ and $f_y$ denote the partial derivatives of
$f(x,y,t)$ in Eq.~(\ref{eq:PFP3}) with respect to the variables $x$ 
and $y$, respectively. They are given by
\begin{equation}
f_x (x,t) =  - 4 \pi^2 (\Omega^2 - A \cos 2 \pi t) \cos 2 \pi x,\,\,
f_y = - 2 \pi \beta \Omega. 
\end{equation}
Note that $J$ is a $2 \times 2$ $q$-periodic matrix.
Let $W(t)=(w^1(t),w^2(t))$ be a fundamental solution matrix  with
$W(0) = I$. Here $w^1(t)$ and $w^2(t)$ are two independent solutions
expressed in column vector forms, and $I$ is the
$2 \times 2$ unit matrix. Then a general solution of the $q$-periodic
system has the following form
\begin{equation}
 \left( \begin{array}{c}
         u(t)  \\
         v(t)
      \end{array}
      \right)
      = W(t)
      \left( \begin{array}{c}
         u(0)  \\
         v(0)
      \end{array}
      \right).
\label{eq:FSM}
\end{equation}
Substitution of Eq.~(\ref{eq:FSM}) into Eq.~(\ref{eq:LE1}) leads to
an initial-value problem to determine $W(t)$
\begin{equation}
{\dot W(t)} = J(t) W(t), \;\;W(0)=I.
\label{eq:FSMEQ}
\end{equation}
It is  clear from  Eq.~(\ref{eq:FSM})  that  $W(q)$   is  just  the
linearized-map matrix   $DP^q(z_0)$.   Hence   the  matrix   $DP^q$
can be obtained through integration of Eq.~(\ref{eq:FSMEQ}) over the
period $q$.

The characteristic equation of the linearized-map matrix
$M (\equiv DP^q)$ is
\begin{equation}
\lambda^2 - {\rm tr}M \, \lambda + {\rm det} \, M = 0,
\end{equation}
where ${\rm tr}M$ and ${\rm det}M$ denote the trace and determinant 
of $M$, respectively. The eigenvalues, $\lambda_1$ and $\lambda_2$,
of $M$ are called the Floquet (stability) multipliers. As shown in
\cite{Lefschetz2}, ${\rm det}\,M$ is calculated from a formula
\begin{equation}
{\rm det}\,M = e^{\int_0^q {\rm tr}\,J dt}.
\label{eq:Det}
\end{equation}
Substituting the trace  of $J$ 
(i.e., ${\rm  tr} J = -  2 \pi \beta \Omega$) into 
Eq.~(\ref{eq:Det}), we obtain an exact analytic result
\begin{equation}
{\rm det}\,M = e^{-2 \pi \beta \Omega q}.
\end{equation}
(Note that ${\rm det} M$ is a constant, independently of the orbits.)
Accordingly, the pair of Floquet multipliers of a periodic 
orbit with period $q$ lies either on the circle of radius 
$e^{-\pi \beta \Omega q}$ or on the  real axis in the complex plane. 
The periodic orbit is stable only when both Floquet multipliers lie 
inside the unit circle. We first note that they never cross the unit 
circle, except at the real axis and hence Hopf bifurcations do not 
occur. Consequently, a stable periodic orbit can lose its stability 
when a Floquet multiplier decreases (increases) through $-1$ $(1)$ on 
the real axis; conversely, an unstable periodic orbit can gain its
stability when a Floquet multiplier increases (decreases) through
$-1$ $(1)$ on the real axis.

When a Floquet multiplier $\lambda$ decreases through $-1$, the 
stable periodic orbit loses its stability via period-doubling 
bifurcation. On the other hand, when a Floquet multiplier $\lambda$ 
increases through $1$, it becomes unstable via pitchfork bifurcation. 
For each case of the period-doubling and pitchfork bifurcations, two 
types of supercritical and subcritical bifurcations occur. For the 
supercritical case of the period-doubling and pitchfork bifurcations, 
the stable periodic orbit loses its stability and gives rise to the 
birth of a new stable period-doubled orbit and a pair of new stable 
orbits with the same period, respectively. On the other hand, for 
the subcritical case of the period-doubling and pitchfork 
bifurcations, the stable periodic orbit becomes unstable by absorbing 
an unstable period-doubled orbit and a pair of unstable orbits with 
the same period, respectively. Hereafter, all these bifurcations, 
associated with instability of a stable periodic orbit, will be called 
the ``normal'' bifurcations. We also note that reverse processes 
of the normal bifurcations can occur for the case of unstable orbits.
That is, when a Floquet multiplier of an unstable orbit increases 
(decreases) through $-1$ $(1)$, it becomes stabilized via ``reverse''
period-doubling (pitchfork) bifurcations. For example, for the reverse 
subcritical period-doubling and pitchfork bifurcations, the unstable 
orbit gains its stability by emitting an unstable period-doubled orbit 
and a pair of unstable orbits with the same period, respectively.
For more details, refer to Ref.~\cite{Guckenheimer}.

We now discuss the Lyapunov exponent and the winding number of an 
orbit in the Poincar{\'{e}} map $P$. Expressing the linearized 
equations $(\ref{eq:LE1})$ for the displacements in terms of the polar 
coordinates $u = r \cos \phi$ and $v = r \sin \phi$,
we have
\begin{equation}
{\dot r} = r [(1+f_x) \sin \phi \cos \phi + f_y \sin^2 \phi], \,\,
{\dot \phi} = - \sin^2 \phi + (f_x \cos \phi + f_y \sin \phi)
\cos \phi.
\label{eq:LE2}
\end{equation}
The motions of the displacements $(r,\phi)$ contain all the 
information about the nearby orbits. Hence we first obtain
the Poincar{\'{e}} maps of an initial displacement $(r_0,\phi_0)$
by sampling the displacements $(r_m,\phi_m)$ at 
the discrete time $t=m$ $(m=1,2,3,\dots)$.
Then the average exponential rate of growth of the radius $r$,
\begin{equation}
\sigma = \lim_{m \rightarrow \infty}
   {1 \over m}  \ln {r_m \over r_0},
\label{eq:Lexp}
\end{equation}
gives the largest Lyapunov exponent $\sigma$, characterizing the 
average exponential rate of divergence of the nearby orbits.
If $\sigma$ is positive, then the orbit is called a chaotic orbit;
otherwise, it is called a regular orbit.
On the other hand, the average rate of increase of the angle $\phi$
(normalized by the factor $2 \pi$),
\begin{equation}
  w = \lim_{m \rightarrow \infty}
    { { |\phi_m - \phi_0|} \over { 2 \pi m}},
 \label{eq:WN}
\end{equation}
gives the winding number $w$, characterizing the average rotation
number of the nearby orbits during the time $1$ (i.e., one iteration 
of $P$). For more details on the Lyapunov exponent and the winding
number, refer to Ref.~\cite{Parlitz}.

\section{Bifurcations of the inverted state and transitions to chaos}
\label{sec:RES}

In this section, by varying the two parameters $A$ and $\Omega$, we 
study bifurcations associated with stability of the inverted state
for a damped case of $\beta=0.2$. It is well known from the theory of
the Mathieu equation \cite{ME} that there exist an infinity of
disconnected stability regions in the $\Omega-A$ plane (i.e., an 
infinity of alternating stable and unstable $A$ ranges exist for any 
given $\Omega$). Consequently, as $A$ is increased, the inverted 
state undergoes a cascade of resurrections (i.e., it stabilizes after
its instability, destabilizes again, and so forth 
{\it ad infinitum}) for any given $\Omega$. We make a detailed 
numerical investigation of bifurcation behaviors associated with such 
resurrections of the inverted state.

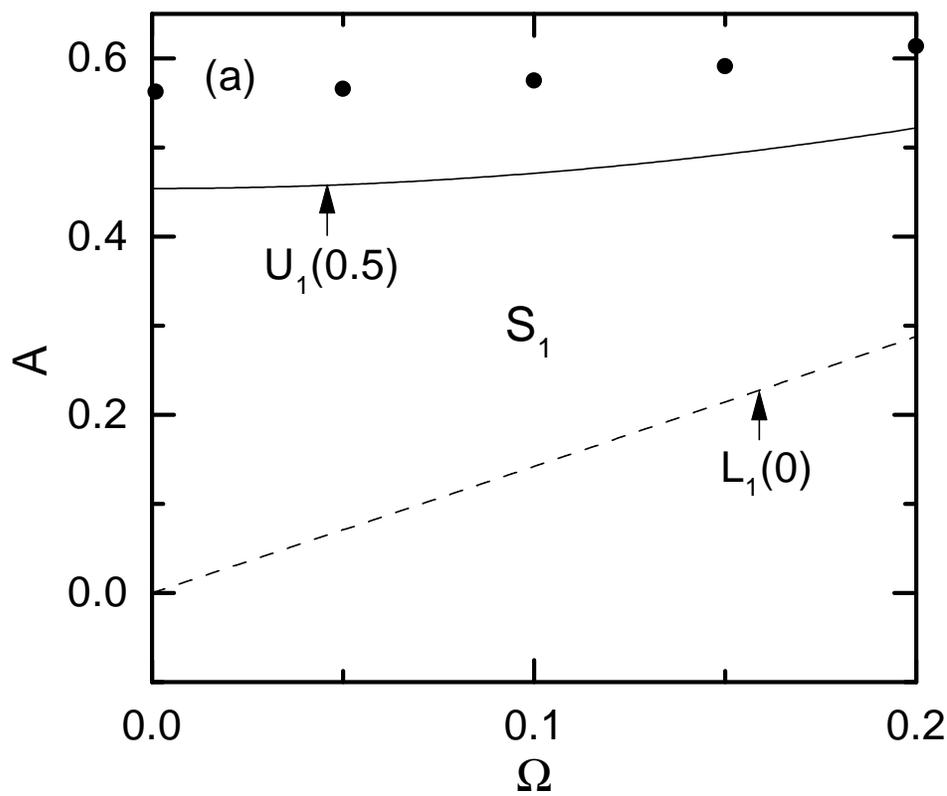
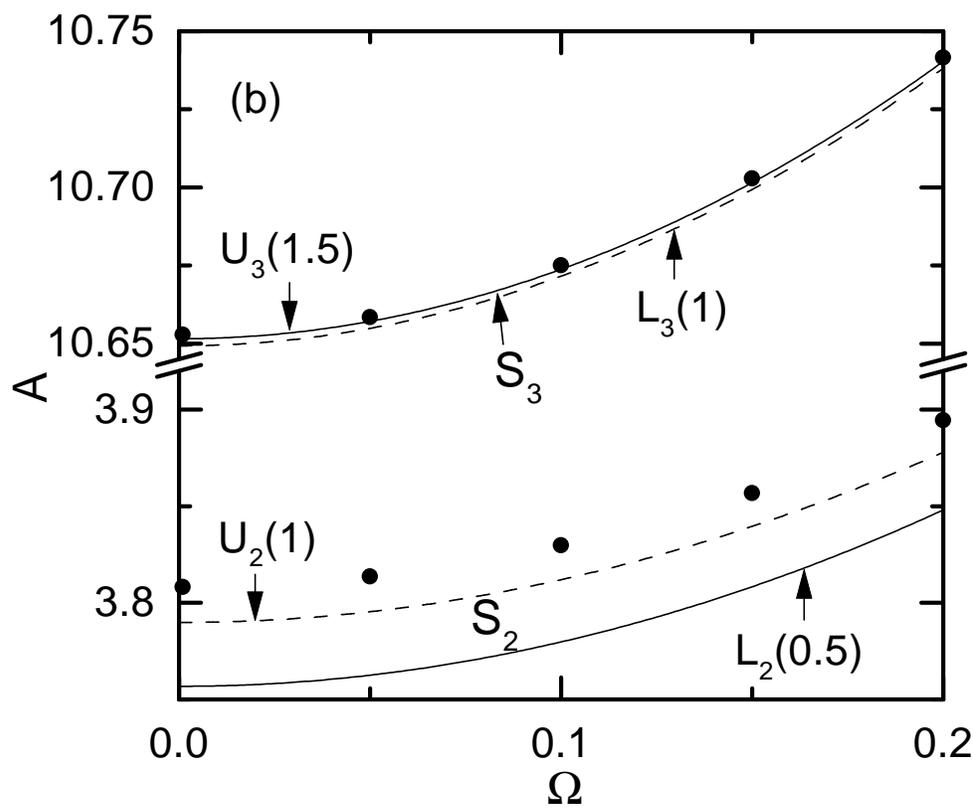
\begin{figure}
\caption{Stability  diagram of  the inverted state in the
parametrically forced pendulum.  The first three stability regions of
the inverted state, denoted by $S_n$ $(n=1,2,3)$, are shown in (a) 
and (b). For each $S_n$, a period-doubling bifurcation occurs on the 
solid boundary curve, while a pitchfork bifurcation takes place on 
the dashed boundary  curve. The lower and upper bifurcation curves, 
$L_n$ and $U_n$, are also labelled by the winding numbers $\omega$ 
of the inverted state as $L_n(\omega)$ and $U_n(\omega)$, 
respectively. The accumulation points of an infinite sequence of 
period-doubling bifurcations, denoted by solid circles, seem to lie 
on the smooth critical lines. For other details, see the text.
     }
\label{fig:SD}
\end{figure}

As explained in Sec.~\ref{sec:SBLW}, the linear stability of a 
periodic orbit with period $q$ in the Poincar\'{e} map $P$ is 
determined from the linearized-map matrix $M(\equiv DP^q)$ of $P^q$. 
The matrix $M$ can be obtained through numerical integration of 
Eq.~(\ref{eq:FSMEQ}) over the period $q$, and then its eigenvalues 
give the Floquet multipliers of the periodic orbit. In such a way, we 
determine the stability regions of the inverted state in the 
$\Omega-A$ plane through numerical calculations of its Floquet 
multipliers $\lambda$'s. The first three stability regions, denoted
by $S_n$ $(n=1,2,3)$, are shown in Fig.~\ref{fig:SD}. Each $S_n$ is 
bounded by its lower stabilization curve, denoted by $L_n$, and by its 
upper destabilization curve, denoted by $U_n$. As the order $n$ is 
increased, the stability region $S_n$ becomes smaller. 

We investigate bifurcation behaviors associated with the 
stabilizations and destabilizations of the inverted state at the 
stability boundary curves in detail. It is found that they depend on 
whether the order $n$ of the stability region $S_n$ is odd or even. 
At the stabilization curves $L_n$ of odd (even) $n$, the unstable 
inverted state becomes stable via reverse pitchfork 
(period-doubling) bifurcations. However, the situation becomes 
reverse for the case of the destabilization curves $U_n$. That is,
the stabilized inverted state loses its stability through normal 
period-doubling (pitchfork) bifurcations when the destabilization 
curves of odd (even) $n$ are crossed. These period-doubling and 
pitchfork bifurcation curves are denoted by the solid and dashed 
curves in Fig.~\ref{fig:SD}, respectively. There are two types of 
supercritical and subcritical bifurcations for the case of the 
period-doubling and pitchfork bifurcations, as explained in 
Sec.~\ref{sec:SBLW}. All the stabilizations occur via the
subcritical bifurcations, while all the destabilizations take place
through the supercritical bifurcations. Consequently, with increasing 
$A$ the inverted state stabilizes via alternating reverse subcritical
pitchfork and period-doubling bifurcations, while it destabilizes
through alternating normal supercritical period-doubling and
pitchfork bifurcations. After each destabilization of the inverted 
state, an infinite sequence of period-doubling bifurcations, leading 
to chaos, also follows and ends at a finite accumulation point. We 
obtain such accumulation points for several values of $\Omega$ 
$(\Omega=0.001,\,0.05,\,0.1,\,0.15,\,0,2)$. They are denoted by solid 
circles in Fig.~\ref{fig:SD} and seem to lie on smooth critical lines.
When crossing these critical lines, period-doubling transitions to
chaos occur.

The lower and upper bifurcation curves, $L_n$ and $U_n$, in 
Fig.~\ref{fig:SD} are also labelled by the winding numbers $\omega$ of 
the inverted state as $L_n(\omega)$ and $U_n(\omega)$, respectively. 
We obtain the winding number $\omega$ of the inverted state through 
numerical integration of the linearized equation $(\ref{eq:LE2})$ over 
the period $1$. It is known that for the pitchfork bifurcations, the 
winding numbers of the inverted state become integers, while they are 
odd multiples of $1/2$ for the period-doubling bifurcations 
\cite{Parlitz}. Note that the winding number $\omega$ of the inverted 
state increases with respect to the order $n$ of the bifurcation 
curves.

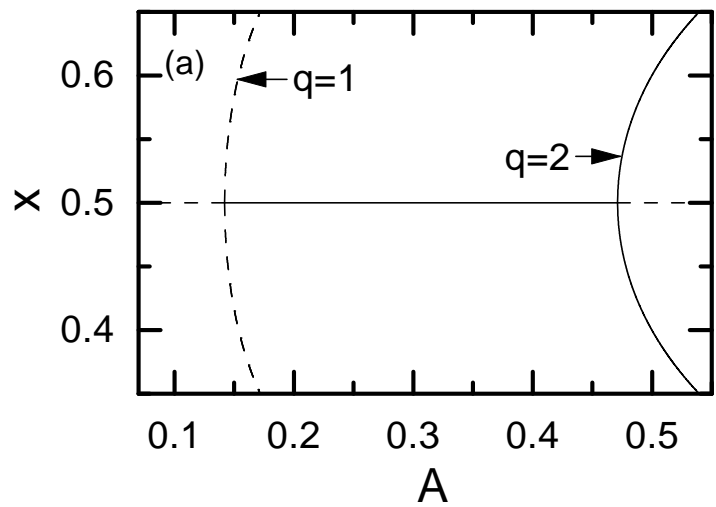
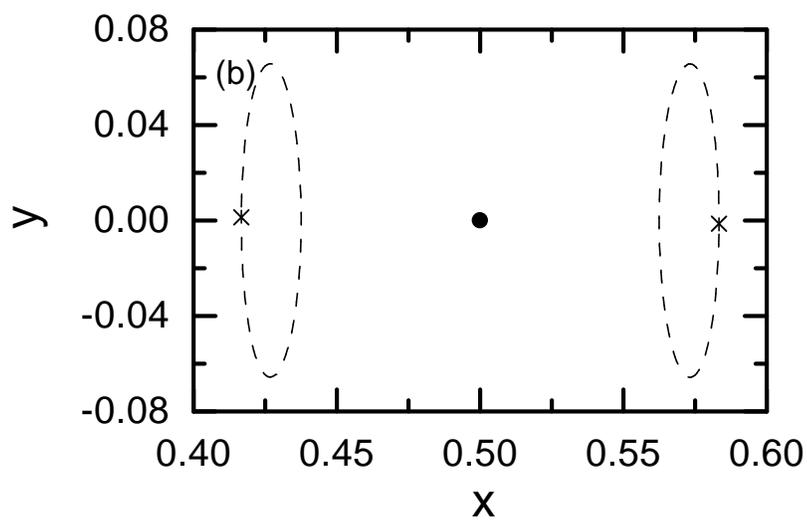
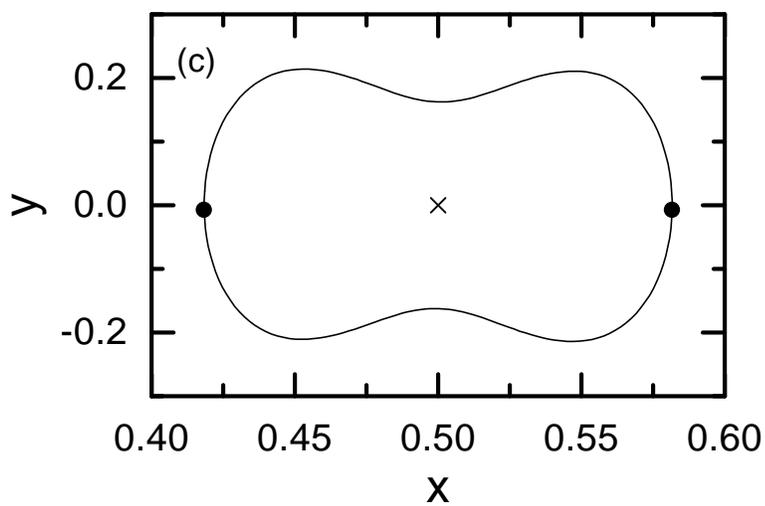
\begin{figure}
\caption{(a) Bifurcation diagram (plot of $x$ versus $A$) in the 
 vicinity of the first resurrection of the inverted state. 
 Here the solid and dashed lines denote stable and unstable orbits, 
 respectively, and $q$ denotes the period of an orbit.
 (b) Phase portraits for $A=0.15$. The stabilized inverted state 
 is denoted by the solid circle. On the other hand, the  phase 
 flows of a conjuate pair of unstable orbits with period $1$
 born via reverse subcritical pitchfork bifurcation are denoted by 
 dashed curves, and their Poincar\'{e} maps are represented by the 
 crosses. (c) Phase portraits for $A=0.5$. The destabilized inverted 
 state is denoted by the cross. On the other hand, the phase flow of 
 a stable orbit with period $2$ born via normal supercritical 
 period-doubling bifurcation is denoted by a solid curve, and its 
 Poincar\'{e} maps are represented by the solid circles.
     }
\label{fig:RES1}
\end{figure}

We now present the concrete examples of bifurcations associated 
with the resurrections of the inverted state for the case $\Omega=0.1$. 
The bifurcation diagrams and the phase-flow and Poincar\'{e}-map plots
are also given for clear presentation of the bifurcations. We first
investigate the bifurcation behaviors associated with the first 
resurrection of the inverted state with increasing $A$. For the 
unforced case of $A=0$, the inverted state is clearly unstable. 
However, when the lower stabilization curve $L_1$ of the first 
stability region $S_1$ is crossed at the first stabilization point 
$A_s(1)=0.142\,066\, \dots$, the unstable inverted state becomes 
stabilized via reverse subcritical pitchfork bifurcation. The 
bifurcation diagram near the first resurrection of the inverted state 
is shown in Fig.~\ref{fig:RES1}(a). Through the reverse subcritical 
pitchfork bifurcation, a conjugate pair of unstable asymmetric orbits 
with period $1$ appears, and their phase portraits for $A=0.15$ are 
shown in Fig.~\ref{fig:RES1}(b). However, when the upper 
destabilization curve $U_1$ of $S_1$ is crossed at the first 
destabilization point $A_d(1)=0.471\,156\, \dots$, the stabilized
inverted state loses its stability via normal supercritical 
period-doubling bifurcation. Consequently, a stable period-doubled 
symmetric orbit appears and its phase portrait for $A=0.5$ is also 
shown in Fig.~\ref{fig:RES1}(c). Note that the winding number $\omega$ 
of the inverted state increases from $0$ to $1/2$, as $A$ is changed 
from $A_s(1)$ to $A_d(1)$. We also study the subsequent bifurcations
with increasing $A$ further. Unlike the case of the inverted state, 
the symmetric $2$-periodic orbit becomes unstable by a 
symmetry-breaking pitchfork bifurcation, which leads to the birth of 
a conjugate pair of stable asymmetric orbits with period $2$. Then 
each $2$-periodic orbit with broken symmetry undergoes an infinite 
sequence of period-doubling bifurcations, ending at its accumulation 
point $A^*_1$ $(=0.575\,154\, \dots)$. Consequently, period-doubling
transition to chaos occurs when the parameter $A$ increases through
$A^*_1$. The critical scaling behaviors of period doublings near the 
critical point $A^*_1$ are the same as those for the 1D maps, as 
will be seen in Sec.~\ref{sec:CB}.

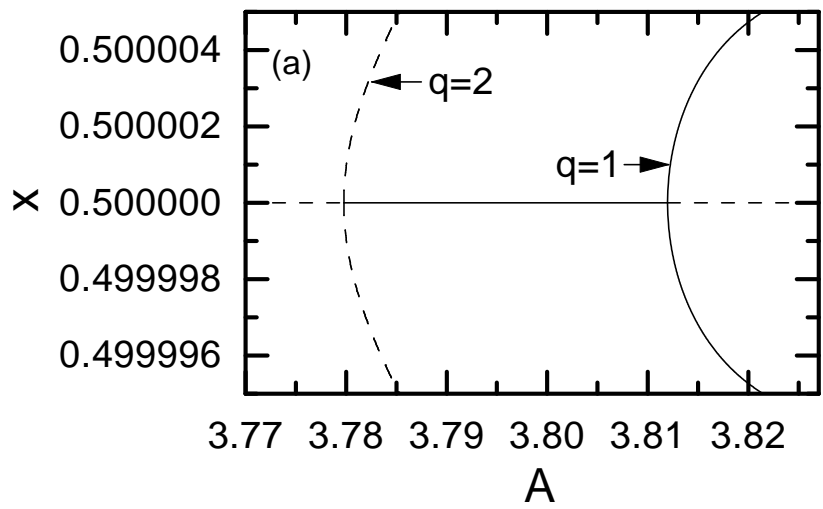
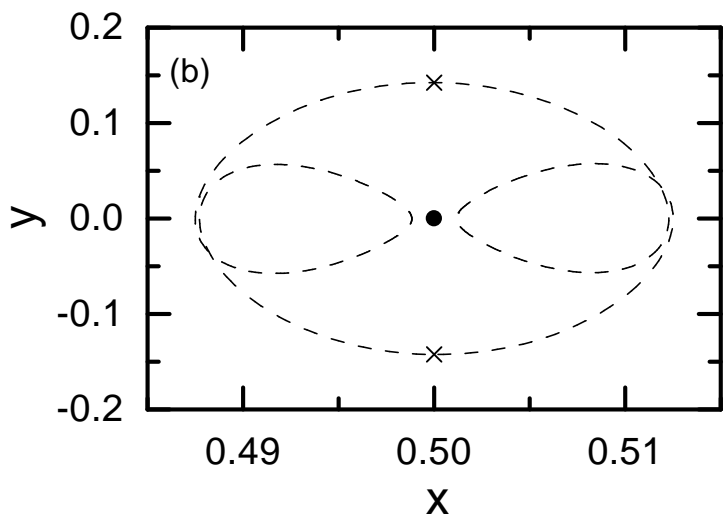
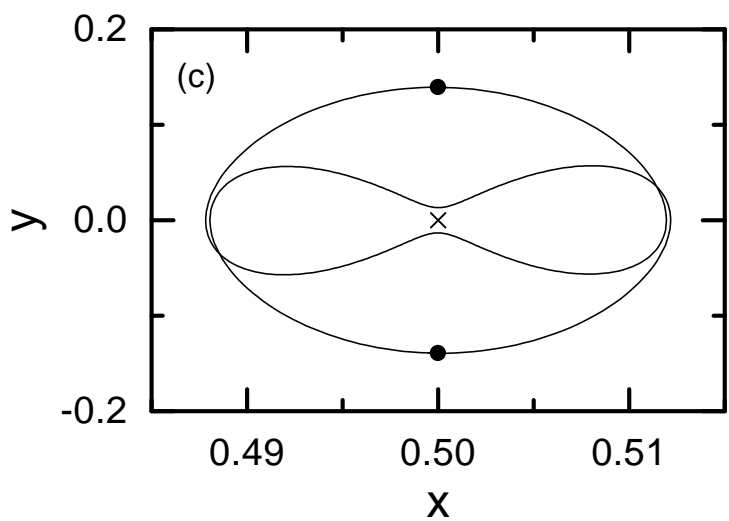
\begin{figure}
\caption{(a) Bifurcation diagram near the second resurrection of the 
 inverted state. The solid and dashed lines and $q$ represent the same 
 as those as in Fig.~1.
 (b) Phase portraits for $A=3.783$. The stabilized inverted state 
 is denoted by the solid circle. On the other hand, the phase flow 
 of an unstable orbit with period $2$ born via reverse subcritical 
 period-doubling bifurcation is denoted by a dashed curve, and its 
 Poincar\'{e} maps are represented by the crosses.
 (c) Phase portraits for $A=3.815$. The destabilized inverted state is
 designated by the cross. On the other hand, the phase flows of a
 conjugate pair of stable asymmetric orbits with period $1$ born via 
 normal supercritical pitchfork bifurcation are denoted by solid 
 curves, and their Poincar\'{e} maps are represented by the solid 
 circles.
     }
\label{fig:RES2}
\end{figure}

With further increase of $A$, we also study the bifurcations
associated with the second resurrection of the inverted state. 
Since the order of $S_2$ is even, the types of 
bifurcations associated with the stabilization and
destabilization become different from those for the case of the
first resurrection. When the lower stabilization curve $L_2$ of $S_2$ 
is crossed at the second stabilization point 
$A_s(2)$ $(=3.779\,771\, \dots)$, a reverse subcritical 
period-doubling bifurcation occurs, which is in contrast to the case 
of the first resurrection. Consequently, the unstable inverted state 
stabilizes with the birth of a new unstable symmetric orbit with 
period $2$. Figure \ref{fig:RES2}(a) shows the bifurcation diagram in 
the vicinity of this second resurrection. The phase portrait of the 
newly-born unstable symmetric $2$-periodic orbit for $A=3.783$ is 
also shown in Fig.~\ref{fig:RES2}(b). However, when the upper 
destabilization curve $U_2$ of $S_2$ is crossed at the second 
destabilization point $A_d(2)=3.811\,973\, \dots$, a normal 
supercritical pitchfork bifurcation occurs, which is also in contrast 
to the case of the first destabilization. As a result, the stable
inverted state destabilizes with the birth of a conjugate pair of 
stable asymmetric orbits of period $1$. The phase portraits of the 
newly-born stable asymmetric orbits with period $1$ for $A=3.815$ are 
shown in Fig.~\ref{fig:RES2}(c). Note that the winding number 
$\omega$ also increases from $1/2$ to $1$, as $A$ is varied from 
$A_s(2)$ to $A_d(2)$. As $A$ is further increased, a second infinite 
sequence of period-doubling bifurcations, leading to chaos, also 
follows and ends at its accumulation point $A^*_2$ 
$(=3.829\, 784\, \dots)$. The critical scaling behaviors of period 
doublings near $A=A^*_2$ are the same as those near the first 
accumulation point $A^*_1$.

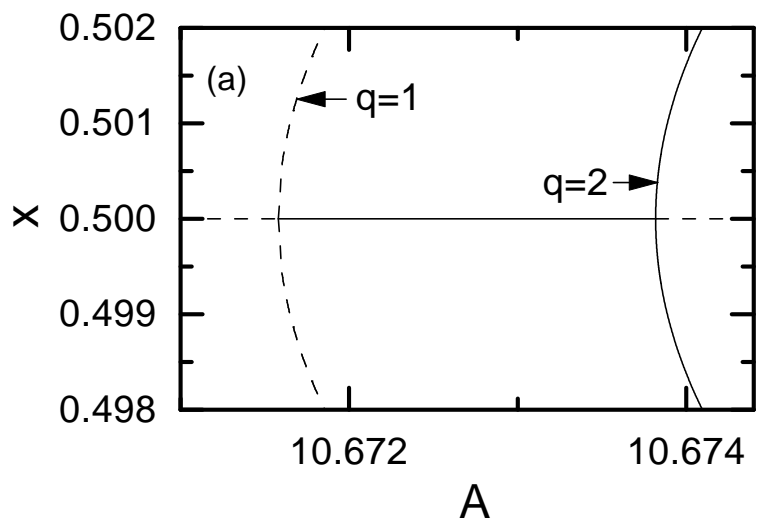
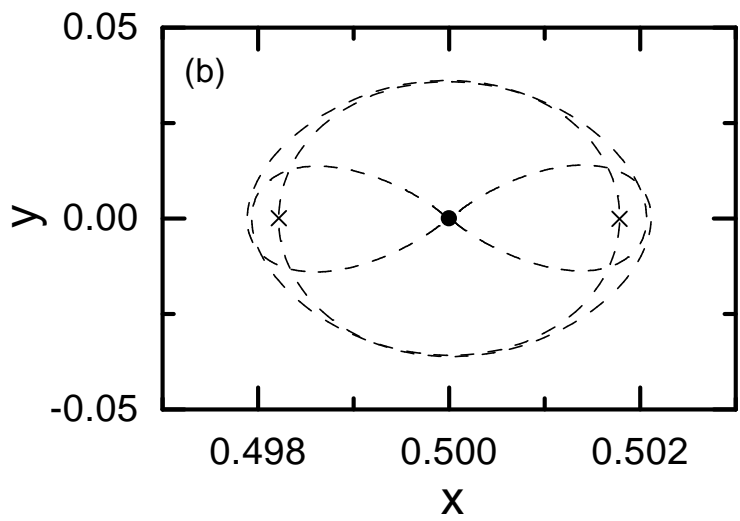
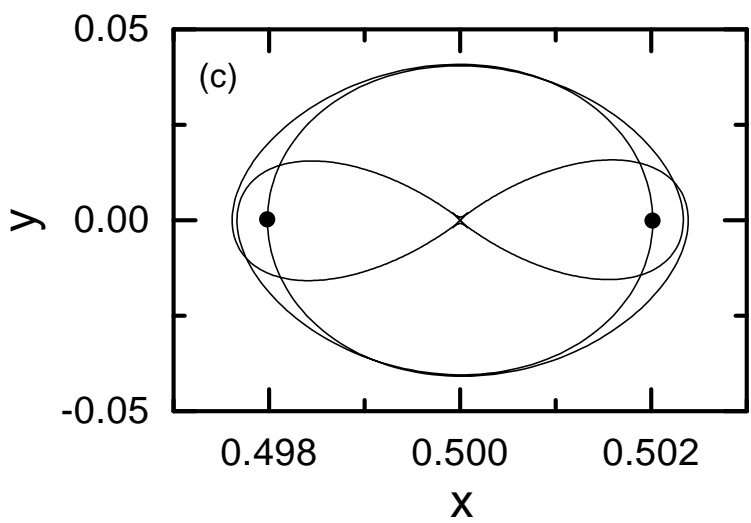
\begin{figure}
\caption{(a) Bifurcation diagram near the third resurrection of the 
 inverted state. The solid and dashed lines and $q$ denote the same as
 those in Fig.~1. Note that the bifurcation behaviors associated with
 the stabilization and destabilization of the inverted state are the 
 same as in Fig.~1. The phase portraits of the orbits associated with 
 the stabilization and destabilization are shown in (b) and (c) for
 $A=10.6718$ and $10.6741$, respectively. For other details see the
 text.
     }
\label{fig:RES3}
\end{figure}

Finally, we investigate the bifurcations associated with the 
third resurrection of the inverted state.
The types of bifurcations associated with the stabilization and
destabilization become the same as those for the case of the
first resurrection, because the order of $S_3$ is odd. As shown in
Fig.~\ref{fig:RES3}(a), the unstable inverted state stabilizes with 
the birth of a conjugate pair of unstable asymmetric orbits with 
period $1$ via reverse subcritical pitchfork bifurcation at the 
third stabilization point $A_s(3)$ $(=10.671\,579\, \dots)$ on the 
lower stabilization curve $L_3$. As $A$ is increased, the stable
inverted state also destabilizes with the birth of a stable 
period-doubled symmetric orbit through a normal supercritical 
period-doubling bifurcation at the third destabilization
point $A_d(3)$ $(=10.673\,818\, \dots)$ on the upper destabilization 
curve $U_3$. The phase portraits of the newly-born orbits for the 
cases of the stabilization and destabilization are shown in 
Figs.~\ref{fig:RES3}(b) and (c) for $A=10.6718$ and $10.6741$, 
respectively. We also note that, as $A$ is changed from $A_s(3)$ to 
$A_d(3)$, the winding number $\omega$ increases from $1$ to $3/2$. 
Since the winding number of the inverted state increases, its nearby 
orbits have an increasing number of loops [e.g., compare 
Figs.~\ref{fig:RES3}(b) and (c) with Figs.~\ref{fig:RES1}(b) and (c)]. 
The subsequent bifurcation behaviors are also the same as those
for the above first case. Consequently, a third infinite sequence
of period-doubling bifurcations, leading to chaos, follows and ends
at its accumulation point $A^*_3$ $(=10.675\, 090\, \dots)$. The
critical scaling behaviors of period doublings are also the same as
those near $A^*_1$.

\section{Critical Behaviors in the Period-doubling Cascades}
\label{sec:CB}

In this section, we first investigate the winding-number sequence of 
the period-doubling cascade and find that the winding numbers at the
period-doubling bifurcation points constitute an alternating sequence
converging to a limit value, as in other oscillaotrs \cite{Parlitz}. 
The orbital scaling behavior and the power spectra of the periodic 
orbits born via period-doubling bifurcations as well as the parameter 
scaling behaviors are then investigated. These critical scaling 
behaviors for all cases studied are found to be the same as those of 
the 1D maps \cite{Feigenbaum}.

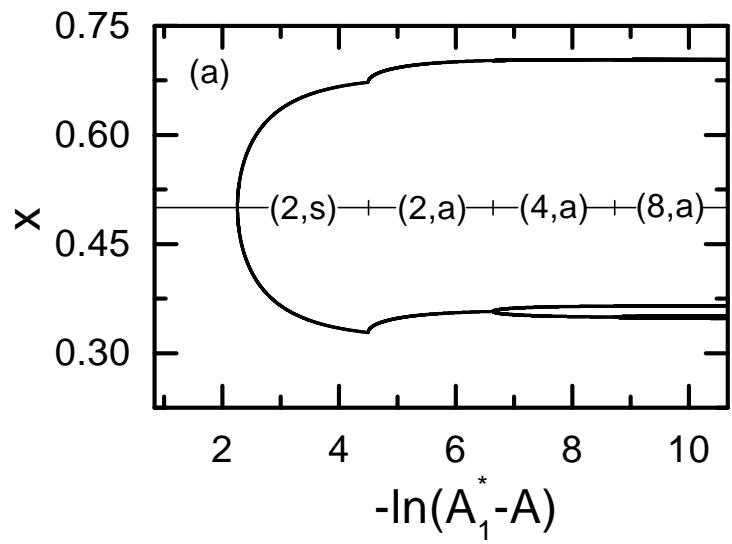
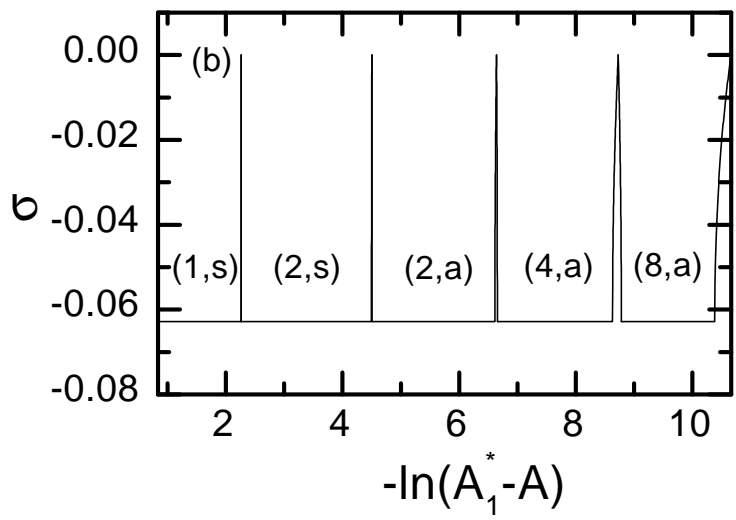
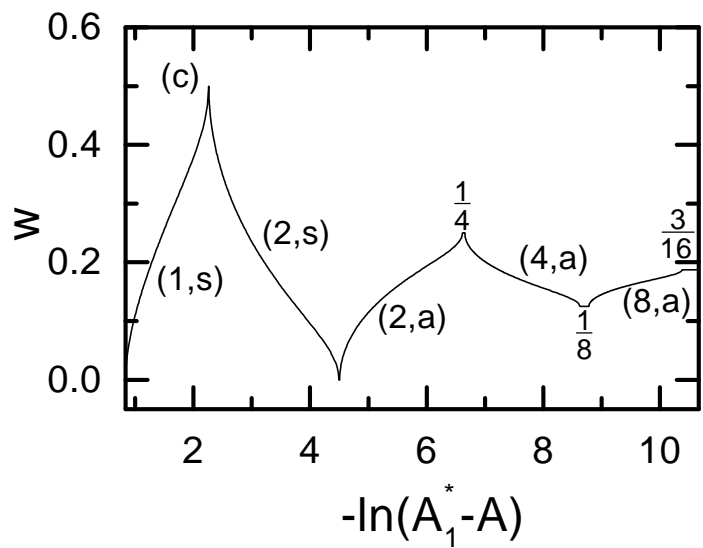
\begin{figure}
\caption{(a) Bifurcation diagram [plot of x versus $-\ln(A^*_1 - A)$] 
of the first period-doubling cascade,
 (b) plot of the largest Lyapunov exponent $\sigma$ versus 
 $-\ln(A^*_1 - A)$, and (c) plot of the winding number $\omega$ 
 versus $-\ln(A^*_1 - A)$ for $\Omega=0.1$. Here $(q,s$ or $a)$ 
 denotes the stable $A$ range of the symmetric or asymmetric orbit 
 with period $q$.
     }
\label{fig:BD}
\end{figure}

As an example, we consider the case $\Omega=0.1$. The first three
accumulation points $A^*_i$'s $(i=1,2,3)$ of the period-doubling 
bifurcations are shown in Fig.~\ref{fig:SD}. Only the critical 
behaviors at the first accumulation point $A^*_1$ are given below, 
because the critical behaviors at all the three accumulation points 
are the same. For this first case, we follow the periodic orbits of 
period $2^k$ up to level $k=8$. As explained in Sec.~\ref{sec:RES}, 
the stabilized inverted state loses its stability for $A=A_d(1)$ 
through a normal supercritical period-doubling bifurcation, giving 
rise to the birth of a stable period-doubled symmetric orbit. 
However, this symmetric $2$-periodic orbit becomes unstable via 
symmetry-breaking pitchfork bifurcation, which results in the birth 
of a conjugate pair of asymmetric orbits with period $2$. Then each 
$2$-periodic orbit with broken symmetry undergoes an infinite 
sequence of period-doubling bifurcations, ending at its accumulation 
point $A^*_1$. Therefore, period-doubling transition to chaos takes 
place when the parameter $A$ increases through $A^*_1$,

Figure \ref{fig:BD}(a) shows the bifurcation diagram of the first 
period-doubling cascade. [For the sake of convenience, only one 
asymmetric orbit of period $2$ is shown in Fig.~\ref{fig:BD}(a).] 
The largest Lyapunov exponent $\sigma$ and the 
winding number $\omega$ in the period-doubling cascade are also
given in Figs.~\ref{fig:BD}(b) and (c), respectively. The largest 
Lyapunov exponent has a constant value $(= - \pi \beta \Omega)$ 
when the Floquet multipliers $\lambda$'s lie on the circle of radius 
$ e^{- \pi \beta \Omega q}$ (in the complex plane), while it 
changes smoothly when $\lambda$'s lie on the real axis. The value of 
$\sigma$ becomes zero at each period-doubling 
bifurcation point. Unlike the case of $\sigma$, the winding number 
$\omega$ takes a constant rational value when $\lambda$'s lie on the 
real axis, and hence it can change only when $\lambda$'s lie on the 
circle of radius $e^{- \pi \beta \Omega q}$. Consequently, $\omega(A)$ 
becomes a step-like function. Rational steps appear near the 
period-doubling bifurcation points, as shown in Fig.~\ref{fig:BD}(c). 
The sequence of the winding-number steps is given by 
$\omega_k = {1\over 6} (1 - { {(-1)^k} \over {2^k} })$ 
$(k=0,1,2, \dots)$; the first three values (corresponding to $k=1,2$
and $3$) are given in Fig.~\ref{fig:BD}(c). Note that the winding 
numbers constitute an alternating sequence converging to its limit 
value $\omega^{(1)}_\infty$ $(=1/6)$. Consequently, the 
quasi-periodic attractor at the first accumulation point $A^*_1$ has 
the winding number $\omega^{(1)}_\infty$. We also study the 
winding-number sequences in the second and third period-doubling
cascades. They are given by
$\omega_k = {1\over 3} (2 + { {(-1)^k} \over {2^k} })$ and
$\omega_k = {1\over 6} (7 - { {(-1)^k} \over {2^k} })$, respectively.
As in the first case, these alternating sequences also converge to 
their limit values  $\omega^{(2)}_\infty$ $(=2/3)$ and 
$\omega^{(3)}_\infty$ $(=7/6)$. Hence, the quasiperiodic
attractors at the second and third accumulation points
$A^*_2$ and $A^*_3$ have their winding numbers  
$\omega^{(2)}_\infty$ and $\omega^{(3)}_\infty$, respectively. Note 
that the winding numbers $\omega^{(i)}_\infty$ of the quasiperiodic
attractors at the accumulation points  $A^*_i$ increase with $i$. 

Table \ref{tab:PSB} gives the $A$
values at which the period-doubling bifurcations occur; at $A_k$,
a Floquet multiplier of an asymmetric orbit with period $2^k$
becomes $-1$. The sequence of
$A_k$ converges geometrically to its limit value $A^*_1$ with
an asymptotic ratio $\delta$:
\begin{equation}
\delta_k = {{A_k - A_{k-1}} \over {A_{k+1} - A_k}} \rightarrow \delta.
\end{equation}
The sequence  of $\delta_k$  is also  listed in Table I. Note
that its limit   value  $\delta$   $(\simeq   4.7)$  agrees  well
with that $(=4.669\cdots)$ for the 1D maps \cite{Feigenbaum}.
We also obtain the value of $A^*_1$ $(=0.575\,154\,232\, \dots)$
by superconverging the sequence of $\{ A_k \}$ \cite{MacKay}.

As in the 1D maps, we are also interested in the orbital scaling 
behavior near the most rarified region. Hence we first locate
the most rarified region by choosing an orbit point $z^{(k)}$
$[=(x^{(k)},y^{(k)})]$ which has the largest distance from its
nearest orbit point $P^{2^{k-1}}(z^{(k)})$ for
$A=A_k$. The two sequences $\{ x^{(k)} \}$ and  $\{ y^{(k)} \}$,
listed in Table \ref{tab:OS}, converge geometrically to their
limit values $x^*$ and $y^*$ with the 1D asymptotic ratio
$\alpha$ $(=-2.502\, \cdots)$, respectively:
\begin{equation}
\alpha_{x,k} = { {x^{(k)} - x^{(k-1)}} \over {x^{(k+1)} - x^{(k)}} }
\rightarrow \alpha, \;\;
\alpha_{y,k} = {{y^{(k)} - y^{(k-1)}} \over
{y^{(k+1)} - y^{(k)}}} \rightarrow \alpha.
\end{equation}
The values of $x^*$ $(=0.350\,686\, \dots)$ and 
$y^*$ $(=0.014\,623\, \dots)$ are also
obtained by superconverging the sequences of $x^{(k)}$ and $y^{(k)}$,
respectively.

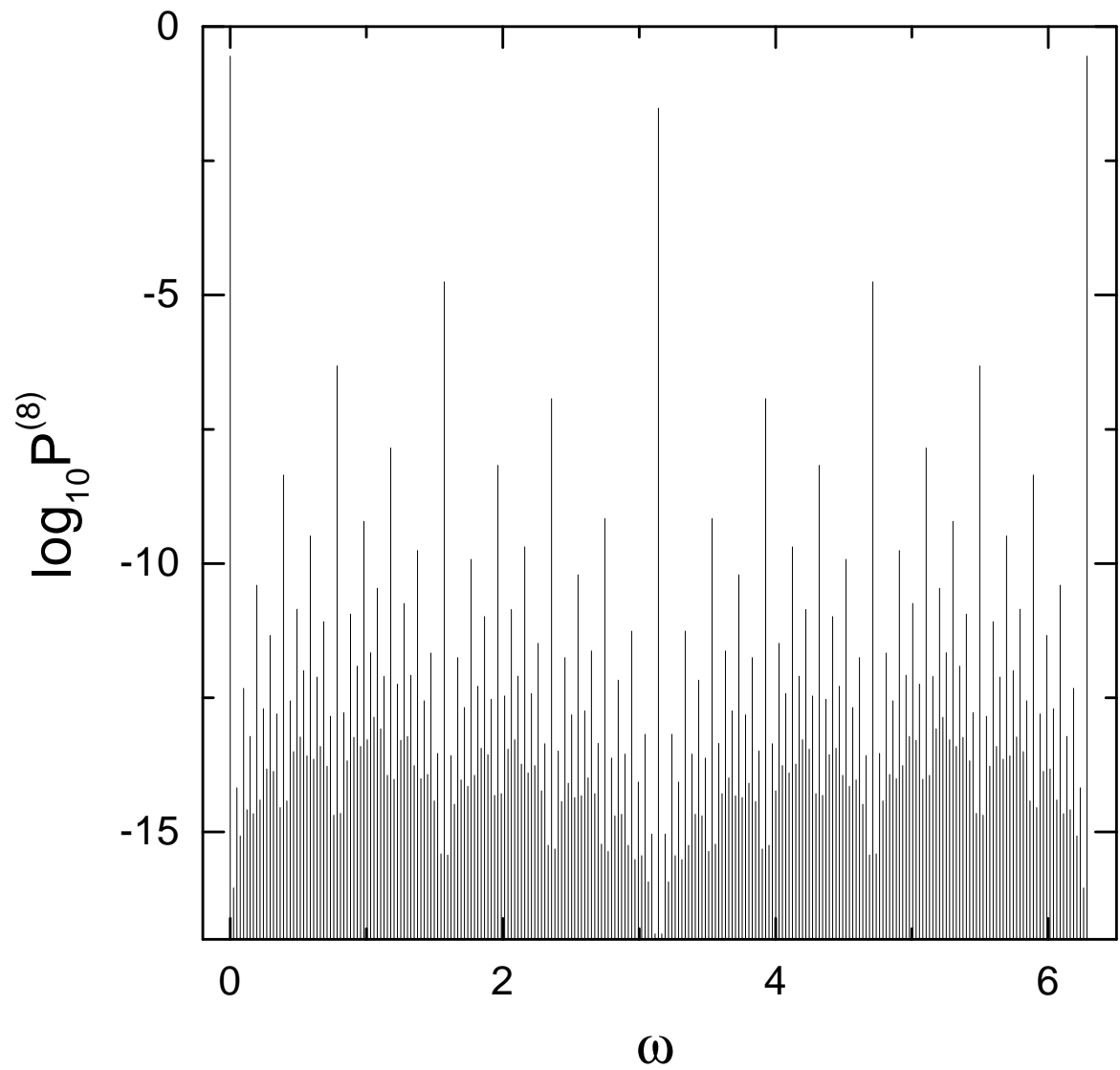
\begin{figure}
\caption{ Power spectrum $P^{(8)}(\omega)$ of level $8$ for
          $A=A_8$ $(=0.575\,154\,223\,350).$
     }
\label{fig:PS}
\end{figure}

We finally study the power spectra of the $2^k$-periodic orbits
at the period-doubling bifurcation points $A_k$.
Consider the orbit of level $k$ whose period is $q=2^k$,
$\{ z^{(k)}_m=(x^{(k)}_m,y^{(k)}_m), \; m=0,1,\ldots,q-1 \}$. Then
its Fourier component of this $2^k$-periodic orbit is given by
\begin{equation}
z^{(k)}(\omega_j) = {1 \over q} \sum_{m=0}^{q-1} z^{(k)}_m e^{-i 
\omega_j m},
\end{equation}
where $\omega_j = 2 \pi j / q$, and $j=0,1,\ldots,q-1$.
The power spectrum $P^{(k)}(\omega_j)$ of level $k$ defined by
\begin{equation}
P^{(k)}(\omega_j) = |z^{(k)}(\omega_j)|^2,
\end{equation}
has discrete peaks at $\omega = \omega_j$.
In the  power spectrum  of the  next $(k+1)$ level, new peaks of the 
$(k+1)$th generation appear at odd harmonics of the fundamental
frequency, $\omega_j = 2 \pi (2j+1) / 2^{(k+1)}$ $(j=0, \ldots,
2^k -1)$. To  classify  the  contributions  of  successive
period-doubling bifurcations  in  the  power spectrum of level $k$, we
write
\begin{equation}
P^{(k)}= P_{00} \delta(\omega) +  \sum_{l=1}^{k}
\sum_{j=0}^{2^{(l-1)}-1}
P^{(k)}_{lj} \delta(\omega- \omega_{lj}),
\end{equation}
where $P^{(k)}_{lj}$ is the height of the $j$th peak of the  
$l$th generation
appearing at $\omega=\omega_{lj}$ $(\equiv 2 \pi (2j+1) / 2^l)$.
As an example, we consider the power spectrum $P^{(8)}(\omega)$ of
level $8$ shown in Fig.~\ref{fig:PS}. The average height of the peaks
of the $l$th generation is given by
\begin{equation}
\phi^{(k)}(l) = {1 \over 2^{(l-1)}} \sum_{j=0}^{2^{l-1}-1} 
P_{lj}^{(k)}.
\end{equation}
It is of interest whether or not the sequence of the ratios of the
successive average heights
\begin{equation}
 2 \beta^{(k)}(l) = \phi^{(k)}(l) / \phi^{(k)}(l+1),
\end{equation}
converges. The  ratios are  listed  in  Table \ref{tab:PSS}. They
seem to approach a limit value, $2 \beta \simeq 21$, which also agrees
well with that $(=20.96 \cdots)$ for the 1D maps \cite{Rudnick}.

\section{Summary}
\label{sec:Sum}
We make a detailed investigation of bifurcations associated with 
resurrections of the inverted state through numerical calculations 
of its Floquet multipliers. It is found that its 
stabilizations occur via alternating reverse subcritical pitchfork 
and period-doubling bifurcations, while its destabilizations take 
place through alternating normal supercritical period-doubling and 
pitchfork bifurcations. An infinite sequence of period-doubling 
bifurcations, leading to chaos, also follows each destabilization of 
the inverted state. The orbital and parameter scaling behaviors near 
the accumulation points $A^*_i$ of the period-doubling cascades are 
also found to be the same as those of the 1D maps, although the 
winding numbers $\omega^{(i)}_\infty$ of the quasi-periodic attractors 
at the accumulation points increase with $i$.

\acknowledgments
Some part of this work has been done while S.Y.K. visited the Centre
for Nonlinear Studies of the Hong Kong Baptist University. This work  
was supported by the Basic Science Research Institute Program, 
Ministry of Education, Korea, Project No. BSRI-97-2401 (S.Y.K.) and
in part by grants from the Hong Kong Research Grants Council (RGC) and 
the Hong Kong Baptist University Faculty Research Grant (FRG).

\begin{table}
\caption{ Asymptotically geometric convergence of the parameter
          sequence $\{ A_k \}$
        }
\label{tab:PSB}
\begin{tabular}{ccc}
$k$ & $A_k$ & $\delta_k$  \\
\tableline
1 & 0.573\,847\,671\,035 &        \\
2 & 0.574\,992\,231\,118 &   8.26 \\
3 & 0.575\,130\,862\,947 &   7.25 \\
4 & 0.575\,149\,971\,647 &   5.68 \\
5 & 0.575\,153\,335\,892&    4.78 \\
6 & 0.575\,154\,039\,822&    4.65 \\
7 & 0.575\,154\,191\,110&    4.69 \\
8 & 0.575\,154\,223\,350&
\end{tabular}
\end{table}

\begin{table}
\caption{Asymptotically geometric convergence of the orbital
           sequences
          $\{ x^{(k)} \}$ and $\{ y^{(k)} \}$.
        }
\label{tab:OS}
\begin{tabular}{cccccc}
$k$ & $x^{(k)}$ & $\alpha_{x,k}$ & $y^{(k)}$ & $\alpha_{y,k}$   \\
\tableline
1 & 0.356\,951\,938 &          &  0.014\,184\,290 &         \\
2 & 0.349\,165\,197 &   -4.016 &  0.014\,680\,027 &  -6.189 \\
3 & 0.351\,104\,216 &   -3.547 &  0.014\,599\,925 &  -2.709 \\
4 & 0.350\,557\,529 &   -3.078 &  0.014\,629\,491 &  -3.543 \\
5 & 0.350\,735\,119 &   -2.608 &  0.014\,621\,145 &  -2.409 \\
6 & 0.350\,667\,034 &   -2.509 &  0.014\,624\,610 &  -2.615 \\
7 & 0.350\,694\,173 &   -2.502 &  0.014\,623\,285 &  -2.448 \\
8 & 0.350\,683\,324 &          &  0.014\,623\,826 &
\end{tabular}
\end{table}

\begin{table}
\caption{ Sequence $2\beta^{(k)}(l)$
$[\equiv \phi^{(k)}(l)/\phi^{(k)}(l+1)]$
of the ratios of the successive average heights of the peaks in the
power spectra
        }
\label{tab:PSS}
\begin{tabular}{cccccc}
\multicolumn{1}{c}{$k$} & \multicolumn{4}{c}{$l$} \\
 & $4$ & $5$ & $6$ & $7$  \\
\tableline
6 & 34.7 & 24.6 & & \\
7 & 34.0 & 24.5 & 21.7 & \\
8 & 33.9 & 24.0 & 21.7 & 21.5
\end{tabular}
\end{table}

\begin{references}
\bibitem{Rem} The sign in front of $\Omega^2$ in Eq.~(\ref{eq:DME})
 is different for the normal and inverted states. For the case of the 
 normal state the sign is positive, while it is minus for the case 
 of the inverted state. The sign is important, because it affects the
 stability.
\end{references}
\end{document}